\documentclass[traditabstract]{aa} 
\usepackage{graphicx}

\usepackage{lscape}
\usepackage{amssymb}
\usepackage{amsmath}

\usepackage{txfonts}

\usepackage{booktabs}
\usepackage{subfigure}
\usepackage{placeins}
\usepackage{verbatim}
\usepackage{adjustbox}
\usepackage[toc,page]{appendix} 
\usepackage{tabularx}
\usepackage{lscape}
\usepackage{textcomp}
\usepackage{gensymb}
\usepackage{hyperref}
\usepackage{multicol}
\usepackage[symbol]{footmisc}
\usepackage{tikz}
\usepackage{caption}
\usepackage{xspace}

\usepackage{natbib}

\usepackage{color}
\usepackage{xtab}
\usepackage{orcidlink}

\newcommand{\bd}{\begin{displaymath}}
\newcommand{\ed}{\end{displaymath}}
\newcommand{\be}{\begin{equation}}
\newcommand{\ee}{\end{equation}}
\newcommand{\beaa}{\begin{eqnarray*}}
\newcommand{\eeaa}{\end{eqnarray*}}
\newcommand{\bea}{\begin{eqnarray}}
\newcommand{\eea}{\end{eqnarray}}

\def\HST{{\it HST}\xspace}
\def\MUSE{MUSE\xspace}

\def\GLEE{\textsc{Glee}\xspace}
\def\GLaD{\textsc{Glad}\xspace}

\def\Sw{\texttt{Sw}\xspace}
\def\Ew{\texttt{Ew}\xspace}
\def\Swo{\texttt{Swo}\xspace}
\def\Ewo{\texttt{Ewo}\xspace}
\def\kmsMpc{\xspace\ensuremath{ \text{~km~s}^{-1}\, \text{Mpc}^{-1}}\xspace}

\def\gleetoolspy{\textit{glee}$ \_ $\textit{tools.py} \xspace}

\defcitealias{schuldt24a}{S24}
\defcitealias{grillo16}{G16}
\begin{document}

   \title{A boost in the precision of cluster-mass models:\\ Exploiting the extended surface brightness of the lensed supernova Refsdal host galaxy}

  \titlerunning{Extended image model of the SN Refsdal host}

   \author{{S. Schuldt}\inst{1}\inst{,2,\orcidlink{0000-0003-2497-6334}}
                        \and
    {C.~Grillo}\inst{1}\inst{,2,\orcidlink{0000-0002-5926-7143}}
    \and
    {A.~Acebron}\inst{3}\inst{,2,\orcidlink{0000-0003-3108-9039}}
    \and
    {P.~Bergamini}\inst{1}\inst{,4,\orcidlink{0000-0003-1383-9414}}
    \and
    {A.~Mercurio}\inst{5}\inst{,6}\inst{,7,\orcidlink{0000-0001-9261-7849}} \and
    {P.~Rosati}\inst{8}\inst{,4,\orcidlink{0000-0002-6813-0632}}
    \and
    {S.~H.~Suyu}\inst{9}\inst{,10,\orcidlink{0000-0001-5568-6052}}
    }

        \institute{Dipartimento di Fisica, Universit\`a  degli Studi di Milano, via Celoria 16, I-20133 Milano, Italy\\
        e-mail: \href{mailto:stefan.schuldt@unimi.it}{\tt stefan.schuldt@unimi.it}
        \and
        INAF -- IASF Milano, via A. Corti 12, I-20133 Milano, Italy
        \and
        Instituto de F\'isica de Cantabria (CSIC-UC), Avda.~Los Castros s/n, 39005 Santander, Spain 
        \and
        INAF -- OAS, Osservatorio di Astrofisica e Scienza dello Spazio di Bologna, via Gobetti 93/3, I-40129 Bologna, Italy
        \and
        Università di Salerno, Dipartimento di Fisica ``E.R. Caianiello'', Via Giovanni Paolo II 132, I-84084 Fisciano (SA), Italy
        \and
        INAF -- Osservatorio Astronomico di Capodimonte, Via Moiariello 16, I-80131 Napoli, Italy
        \and
        INFN – Gruppo Collegato di Salerno - Sezione di Napoli,  Dipartimento di Fisica "E.R. Caianiello", Università di Salerno, via Giovanni Paolo II, 132 - I-84084 Fisciano (SA), Italy
        \and
        Dipartimento di Fisica e Scienze della Terra, Università degli Studi di Ferrara, via Saragat 1, 44122 Ferrara, Italy
        \and
        Technical University of Munich, TUM School of Natural Sciences, Department of Physics,  James-Franck-Stra{\ss}e 1, 85748 Garching, Germany
        \and
        Max-Planck-Institut f{\"u}r Astrophysik, Karl-Schwarzschild Stra{\ss}e 1, 85748 Garching, Germany
         }

   \date{Received --; accepted --}

 
  \abstract{Combining deep \textit{Hubble} Space Telescope (\HST) images and extensive data from the Multi-Unit Spectroscopic Explorer, we present new mass models of the cluster MACS J1149.5+2223, strongly lensing the supernova (SN) Refsdal, fully exploiting the source surface-brightness distribution of the SN host for the first time. In detail, we incorporated 77,000 \HST pixels, in addition to the known 106 point-like multiple images, in our modeling. We considered four different models to explore the effect of the relative weighting of the point-like multiple image positions and flux distribution of the SN host on the model optimization. When the SN host's extended image is included, we find that the statistical uncertainties of all 34 free model parameters are reduced by factors ranging from one to two orders of magnitude compared to the statistical uncertainty of the point-like only model, irrespective of the adopted different image weights. We quantified the remarkably increased level of precision with which the cluster's total mass and the predicted time delays of the SN Refsdal multiple image positions can be reconstructed. We also show the delensed image of the SN host, a spiral galaxy at $z_\text{SN}=1.49$,  in multiple \HST bands. In all those applications, we obtain a significant reduction of the statistical uncertainty, which is now below the level of even the small systematic uncertainty on the mass model that could be assessed by the different approaches. These results demonstrate that with extended image models of lensing clusters it is possible to measure the cluster's total mass distribution, the values of the cosmological parameters, and the physical properties of high-redshift sources with an unparalleled precision, making the typically not-quantified systematic uncertainties now crucial.
}


   \maketitle

\section{Introduction}
\label{sec:intro}

Massive objects, such as galaxies and galaxy clusters, can cause the strong gravitational lensing effect; namely, background sources are observed multiple times. Since this effect depends on the total mass of the lens, it is a powerful tool for dark matter \citep[DM; e.g.,][]{grillo15, schuldt19, meneghetti20, wang22} and galaxy evolution \citep[e.g.,][]{annunziatella17, mercurio21} studies. Furthermore, thanks to the lensing magnification, it enables the detection and investigation of otherwise too-faint, high-redshift sources \citep[e.g.,][]{vanzella21, vanzella22, mestric22, mestric23, stiavelli23, morishita24}. 

As proposed by \citet{refsdal64}, the values of the time delays between the multiple images of a strongly lensed supernova (SN) can be used to measure the value of the Hubble constant, $H_0$, which describes the current expansion rate of the Universe. This technique recently gained significant attention, given the current discrepancy between the measurements from the cosmic microwave background from Planck, resulting in $H_0 = \left(67.4 \pm 0.5 \right) \kmsMpc$ \citep{planck20}, and the Cepheid distance ladder approach used by the SH0ES team, providing $H_0 = \left( 73.0 \pm 1.0 \right) \kmsMpc$ \citep{riess22}. The discrepancy raises the question of unknown systematics \citep[e.g.,][]{freedman24, riess24} or new physics \citep[e.g.,][]{divalentino21}, and time-delay cosmography (TDC) could play a major role in its clarification. 

\begin{figure}[t!]
    \centering
    \includegraphics[trim={103 38 90 40},clip, width=\linewidth]{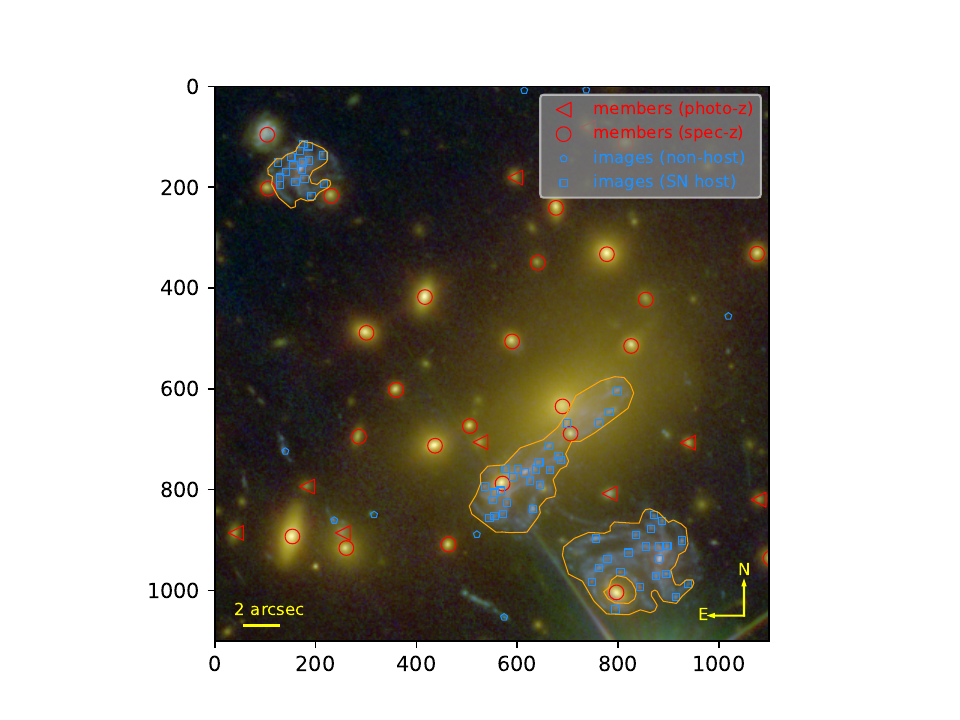}
    \caption{Color-composite image (F435W for blue, F606W + F814W for green, and F105W + F125W + F140W + F160W for red) of the cluster core with the SN Refsdal and its host. Shown are the cluster members (red triangles for photometrically selected members and red circles for spectroscopically confirmed members) and the multiple image systems (blue squares for SN Refsdal and the host and blue pentagons otherwise). We indicate the regions (in orange) used for the SN host reconstruction and extended image model.}
    \label{fig:HFF_zoom}
\end{figure}

The first-discovered spatially resolved gravitationally lensed SN (glSN; \citealt{kelly15, kelly16a}) was imaged multiple times by the massive galaxy cluster MACS~J1149.5+2223 (hereafter MACS1149; see Fig.~\ref{fig:HFF_zoom} for the cluster core), which lies at a redshift of $z_\text{c}=0.5422$ \citep{grillo16}. The glSN, called SN Refsdal and located at $z_\text{SN}=1.489$, was lensed into six multiple images, while one image (called SY) appeared before the discovery, so only four time delays could be measured \citep{rodney21, kelly23b}. These time delays, together with the observed positions of several multiple images of different background sources, were used by \citet{kelly23} to measure the value of the Hubble constant, resulting in $H_0 = \left( 64.8^{+4.4}_{-4.3} \right) \kmsMpc$, from the combination of eight different mass models, and $H_0 = \left( 66.6^{+4.1}_{-3.3} \right) \kmsMpc$, by combining their two best-fitting models  \citep[Oguri-a$^\star$ and Grillo-g from][]{treu16}. By directly including the measured time delays in the frozen Grillo-g mass model \citep[hereafter \citetalias{grillo16}]{grillo16}, and by assuming a more general background cosmological model, \citet{grillo24} blindly obtained $H_0 = \left( 65.1^{+3.5}_{-3.4} \right) \kmsMpc$, accounting for both the statistical and systematic uncertainties. 

More recently, two other glSNe, SN H0pe, which is strongly lensed by the galaxy cluster PLCK~G165.7+67.0 \citep{frye23_SNH0pe_astronote}, and SN Encore \citep{pierel24_Encore}, located in the same host galaxy as the SN Requiem \citep{rodney21} behind the cluster MACS~J0138.0$‑$2155, were discovered. A first estimate of the Hubble constant using the SN H0pe time delays was presented by \citet{pascale25}, resulting in a (post-blind) value of $H_0 = \left( 71.8 ^{+9.2} _{-8.1} \right) \kmsMpc$, while the analysis with seven different mass models for the SN Encore system resulted in $H_0 = \left( 66.9  ^{+11.2} _{-8.1} \right) \kmsMpc$ \citep{pierel25, suyu25}. 

Although significant effort has been made to find more glSNe and improve the $H_0$ measurements, the relative uncertainty on the value of $H_0$ is still relatively high ($\sim5$--$6\%$ for SN Refsdal and $12\%$ for SN H0pe) when compared with that on the time delays. The uncertainty on the reconstructed total mass distribution of the lens used for TDC contributes, together with that on the measured time delays, to the total error budget of the $H_0$ value. In particular, for SN Refsdal, the cluster total mass model uncertainty ($\sim5\%$) strongly dominates over the time delay uncertainty ($\sim 1\%$). This reflects the necessity of improving the cluster's total mass reconstruction and of better understanding its statistical and systematic uncertainties \citep[e.g.,][]{grillo20}. So far, all the considered cluster total mass models for glSNe rely on the identified multiple image positions of different background sources \citep[e.g.,][]{treu16, kelly23, acebron25, ertl25, pascale25}, which represents the state-of-the-art ``point-like" approach on cluster scale. However, on galaxy scale, where so far only multiply-imaged quasars have been exploited for TDC \citep[e.g.,][]{suyu10b, wong17, birrer19, chen19, shajib20, wong24, tdcosmo25}, modeling the extended surface-brightness distribution of the quasar's host galaxy is common, and this has been shown to provide significantly more information than when just using the quasar's point-like images. Lately, some remarkable effort has been made in the direction of the extended source reconstructions on group-scale systems \citep{ding25_Cassowary19, bolamperti24, wang22}, while so far, solely \citet{acebron24} has successfully pushed forward this extremely time-consuming approach on a cluster-scale system, specifically on the galaxy cluster SDSS~J1029+2623. This cluster contains a strongly lensed quasar with a measured time delay and is therefore well suited for TDC. In this work, we followed a similar approach to \citet{acebron24} and relied on the same modeling software, \GLEE \citep{suyu10a_GLEE, suyu12b_GLEE}, while focusing on weighting schemes and the impact on the statistical uncertainty rather than the extracted source properties. In contrast, the approach and the aim of \citet{sengul23} are notably different, as the authors focused on distortions of giant arcs to detect smaller perturbers and measure their masses (such as DM components or supermassive black holes), did not constrain the total mass of the cluster using a direct reconstruction of the arc as we did here.

\citet[][hereafter \citetalias{schuldt24a}]{schuldt24a} recently published an enhanced strong lens mass model of MACS1149, incorporating new VLT/MUSE observations and following the state-of-the-art point-like strong lens modeling approach. Exploiting the new spectroscopic data, and thanks to a reanalysis of previous observations, this model includes 17 previously unknown multiple images from six different background sources, extending the covered source redshift range from $z=3.703$ (system 14 in \citetalias{grillo16}) to $z=5.983$. The availability of many multiple images at various redshifts gives the significant advantage of reducing the mass-sheet degeneracy, which is one of the main sources of systematic uncertainties in gravitational lensing studies. This newly presented point-like model resulted in a root-mean-square (RMS) value of the distances between the observed and model-predicted positions of the multiple images of 0\farcs39, with only 34 free parameters.

Building upon this point-like mass model, we present the first extended model of the surface brightness of the host galaxy of SN Refsdal, which is so far the best studied glSNe. We incorporated 77,000 \textit{Hubble} Space Telescope (\HST) pixels of the SN host galaxy (i.e., the pixels within the orange contours in Fig.~\ref{fig:HFF_zoom}), which translates into an increase of the number of observables by two orders of magnitude, while the underlying parametrization of the cluster total mass remains the same as in \citetalias{schuldt24a}. This approach also allowed us to reconstruct the delensed surface-brightness distribution of the SN host, which is a spiral galaxy at $z_\text{SN}=1.49$. With fixed cosmology, we carried out a detailed analysis of the effect of this model extension on the statistical errors on the cluster's total mass and on the model-predicted time delays. Moreover, we tested different weighting schemes between the point-like and extended image contributions to the value of the total $\chi^2$. These insights directly expanded the work presented by \citet{acebron24} adopting a single weighting scheme and will pave the way toward a new generation of cluster mass models that will enable more precise measurements of the value of the Hubble constant and of other quantities such as the magnification factors of background sources. In particular, with the ongoing \textit{Euclid} \citep{laureijs11} and Legacy Survey of Space and Time at the \textit{Rubin} Observatory \citep{ivezic08} observational campaigns, the number of detected glSNe will significantly increase \citep[e.g.,][]{arendse24}, and complete lensing analyses for TDC will become even more crucial.

The paper is organized as follows. We introduce the exploited observables and the parametrization of the strong lens mass model in Sect.~\ref{sec:model}, along with a brief discussion of the optimization procedure for an extended source. In Sect.~\ref{sec:results} we show our novel results and compare them with those from our previous point-like model from \citetalias{schuldt24a}. Section~\ref{sec:td_mass} summarizes the impact of the new modeling on the derived quantities, such as the total mass of the cluster and the predicted time delays. We conclude our work in Sect.~\ref{sec:conclusion}.

Throughout this paper, we assume a flat $\Lambda$ cold dark matter ($\Lambda$CDM) cosmology
with $H_0 = 70\kmsMpc$ and $\Omega_\text{M} = 1-\Omega_\text{DE} = 0.3$. In this cosmology, $1\arcsec$ corresponds to 6.36 kpc at the cluster redshift, $z_\text{c} = 0.5422$.

\FloatBarrier
\section{Lens mass model}
\label{sec:model}

The presented mass model was developed with the \GLEE \citep{suyu10a_GLEE, suyu12b_GLEE} modeling software, while we also made use of \gleetoolspy \citep{schuldt23b}. \GLEE allowed us to obtain the total mass distribution of the cluster with a parametric approach, which we describe further in Sect.~\ref{sec:model:parameterization}. The values and errors of the model parameters are typically constrained by the identified multiple image positions of the strongly lensed sources (see Sect.~\ref{sec:model:ImPos}), but \GLEE also allows one to exploit the extended images of lensed sources in the mass model as constraints \citep[e.g.,][]{suyu10a_GLEE, acebron24}. This additionally enables the reconstruction of the (unlensed) surface-brightness distribution of those sources by minimizing the difference between the model-predicted and observed flux values on the regions covered by the multiple images. This procedure is detailed in Sect.~\ref{sec:model:ext}.

\subsection{Mass parametrization}
\label{sec:model:parameterization}

The presented extended-image mass model is based on the point-like mass model of \citetalias{schuldt24a} and assumes the exact same cluster total mass parametrization to allow a fair comparison to the model from \citetalias{schuldt24a}, considered as reference mass model. In short, the overall cluster DM distribution is described by three pseudo-isothermal elliptical mass distribution \citep{kassiola93} profiles, two of which are centered in the cluster core. Each of them is characterized by six parameters: the coordinates of the center, $x_\text{H}$ and $y_\text{H}$; the axis ratio, $q_\text{H}$; its orientation, $\phi_\text{H}$; the halo strength (or Einstein radius), $\theta_\text{H}$; and the core radius, $r_\text{c,H}$. Moreover, we include, as in \citetalias{schuldt24a}, a fourth circular ($q_\text{H} \equiv 1$) halo associated with the galaxy-group halo located in the north of the cluster. These cluster halos describe mostly the DM component of the galaxy cluster, which hereafter we refer to as the DM halo.

Furthermore, we included the 308 securely identified cluster members from \citetalias{schuldt24a}, of which 195 (63\%) are spectroscopically confirmed. Following \citetalias{grillo16}, the mass distribution of two cluster members located close to the SN Refsdal host are described by a dual pseudo-isothermal elliptical (dPIE) mass distribution \citep{eliasdottir07, suyu10a_GLEE} with vanishing core radius. All of the remaining 306 cluster members are represented through a circular ($q\equiv 1$) dPIE profile, whose truncation and Einstein radii are scaled according to their magnitudes in the \HST F160W band. While these assumptions result in an approximation of the cluster's total mass distribution, these mass profiles and scaling relations are commonly adopted for cluster mass models. The cluster member distribution in the cluster core is shown in Fig.~\ref{fig:HFF_zoom} with red triangles if photometrically identified, and red circles if spectroscopically confirmed. The distribution of all cluster members is displayed in Fig.~1 of \citetalias{schuldt24a}.

\subsection{Multiple image positions}
\label{sec:model:ImPos}

While the mass parametrization (see Sect.~\ref{sec:model:parameterization}) was kept the same, we considered four different combinations and weightings of the observables, as summarized in Table~\ref{tab:overview}. In the first two models, hereafter referred to as \Sw and \Ew, we included all multiple images from \citetalias{schuldt24a}, namely 63 multiple images belonging to 18 knots of the SN Refsdal host ($z_\text{SN}=1.489$) and 43 multiple images from 16 other strongly lensed sources, covering the redshift range between 1.24 and 5.98. In the other two models, hereafter referred to as \Ewo and \Swo, we excluded the identified multiple-image positions of the SN Refsdal and of its host\footnote{From here on, we simplify the wording and write images of the SN host when also referring to the SN images itself.}. The observed positions of the SN Refsdal and the identified multiple images of the SN host are shown in Fig.~\ref{fig:HFF_zoom} (blue squares). We refer to \citetalias{grillo16} and \citetalias{schuldt24a} for their exact coordinates and redshifts.

\begin{table*}[ht!]
    \caption{Overview of the considered mass models and their chi-square values for the corresponding best-fit models. We note that only $\chi^2_\text{img,S24,w}$, $\chi^2_\text{ext,orig}$, and RMS (w) can be fairly compared across different models, as they are computed assuming the same observational uncertainties. \label{tab:overview}}
    \begin{center}
    \begin{tabular}{cccc|cccccc}
    \hline \hline 
Model   & SN host & weighting & d.o.f. &$\chi^2_\text{img}$ &$\chi^2_\text{ext}$ & $\chi^2_\text{red}$ &$\chi^2_\text{img,S24,w}$ &$\chi^2_\text{ext,orig}$ & RMS (w) \\
\noalign{\smallskip} \hline \noalign{\smallskip}
S24  & Y & $\chi^2_\text{img}$ only     & 110 & 111.3     & --     & 1.01 & 111.3 & 37,105 & 0\farcs39\\
\Ew  & Y & $\chi^2_\text{ext}$-dominant & 70,272 & 394    & 68,853 & 0.99 & 394   & 30,286 & 0\farcs73\\ 
\Sw  & Y & $\chi^2_\text{ext} \sim \chi^2_\text{img}$     & 73,232 & 35,455 & 36,234 & 0.98 & 138.5 & 33,673 & 0\farcs44\\ 
\Ewo & N & $\chi^2_\text{ext}$-dominant & 70,151 & 357    & 69,198 & 0.97 & 361   & 30,406 & 0\farcs70 \\ 
\Swo & N & $\chi^2_\text{ext} \sim \chi^2_\text{img}$     & 73,371 & 34,803 & 36,212 & 0.99 & 141   & 33,654 & 0\farcs44\\ 
    \noalign{\smallskip} \hline 
    \end{tabular}
    \end{center}
\textbf{Note.} The columns give from left to right the model acronym, whether the SN host image positions are included (Y) or not (N), the $\chi^2$ term weighting, the number of degrees of freedom (d.o.f.), the image position chi-square value, as obtained from the optimization, the chi-square value of the extended image, as obtained from the optimization, the total reduced chi-square defined as $\chi^2$/d.o.f., the image position chi-square, adopting the uncertainties of \citetalias{schuldt24a} and always including the multiple image positions of the SN host, the extended chi-square value, by adopting the original error map as defined in Eq.~\ref{eq:errmap}, and, finally, the root mean square (RMS) from all multiple image positions (to allow for a fair comparison across the five models).
\end{table*}

Since the cluster's total mass distribution is relatively complex and the image multiplicity high, we adopted the expression of \citetalias{schuldt24a} for the image-position $\chi^2$ term, namely
\be
\chi^2_\text{img} = \sum_{j=1}^{N^\text{fam}} \left\{ \begin{array}{cc}
     \sum_{k=1}^{N_j^\text{obs}}  \frac{\left(x_{j,k}^\text{obs} - x_{j,k}^\text{pred} \right)^2 }{\Delta x_{j,k}^2} + \frac{\left(y_{j,k}^\text{obs} - y_{j,k}^\text{pred} \right)^2 }{\Delta y_{j,k}^2} &  \text{if~} N_j^\text{obs} \leq N_j^\text{pred}\\
     \infty & \text{otherwise}
     \label{eq:chi2}
\end{array} \right. ,
\ee
which ensures the correct image multiplicity for all sources. The mass model relying only on the point-like image positions was obtained by minimizing this term, and keeping it here allowed us to draw a direct comparison to the mass model presented by \citetalias{schuldt24a}. In this equation, $N^\text{fam}$ describes the number of families, $N^\text{obs}_j$ and $N^\text{pred}_j$ the number of observed and predicted image positions of family $j$, respectively; $x^\text{obs}$ and $y^\text{obs}$ the observed $x$ and $y$ coordinates, respectively; and $x^\text{pred}$ and $y^\text{pred}$ are the predicted $x$ and $y$ coordinates, while $\Delta$ describes the image position uncertainty of the corresponding image position. In the next section, we provide further details on the parameter optimization. 

\subsection{Extended image model and source surface-brightness reconstruction}
\label{sec:model:ext}

In addition to the image positions described in Sect.~\ref{sec:model:ImPos}, we included the flux values of the pixels of the SN host in our mass model as observables. Since we only wanted to model and reconstruct the surface-brightness distribution of the SN host, we followed the approach of \citetalias{grillo16} and created a difference image between the \HST\ filters F435W and F606W. In detail, we used a representative subset of spectroscopically confirmed cluster members 
to determine a scaling factor for the F606W band to best subtract the light of these cluster members, resulting in an image with only the (lensed) background sources. The best aperture size to estimate the scaling factor was determined in such a way that the remaining fluxes within 1\arcsec around the selected cluster members were minimal. 

To create error maps, which were not directly available for the \HST images, we approximated them for both filters using \gleetoolspy \citep{schuldt23b}. In this approximation, the error for pixel $j$ of filter $k$ is defined as
\be
\sigma^2_{\text{tot,}k,j} = \left\{ \begin{array}{cl} \sigma^2_{\text{poisson,}j}+ \sigma^2_{\text{bkgr,}j} & \text{if} \, \sigma_{\text{bkgr,}j} \leq \sigma_{\text{poisson,}j}\\
\sigma_{\text{bkgr,}j}^2 & \text{otherwise}
\end{array} ,
\right.
\ee
with the Poisson noise \citep{hasinoff12}
\be
\sigma_{\text{poisson,}j} = \sqrt{\frac{ |d_j|}{t_j}} \, ,
\ee
which we obtained from the observed pixel intensity, $d_j$ (in $e^-$-counts per second) and the exposure time, $t_j$,  of pixel $j$, and $\sigma_{\text{bkgr,}j}$ approximated as a constant value equal to the standard deviation from an empty region. After generating both error maps separately, we combined them via
\be
\sigma_{\text{tot,}j} = \sqrt{\sigma_{\text{tot, F435W,}j}^2 + \sigma_{\text{tot, F606W,}j}^2} \, ,
\label{eq:errmap}
\ee

\noindent without different weightings of the individual error maps, as we applied a weighting to the final error map (see Table~\ref{tab:overview}).

In addition, we needed to define which pixels of the difference image belong to the SN host and should be taken into account during the modeling. For this, we created an arcmask \citep[see, e.g.,][for further details]{ertl23}. The considered area is shown with orange contours in Fig.~\ref{fig:HFF_zoom}, and contains $N_\text{pix}$ = 77,000 pixels with a pixel size of 0\farcs03.

In each iteration of the parameter optimization, all pixels within the arcmask are mapped from the observed image plane onto the source plane, using the given mass model through the ray-tracing equation. In this way, the source was reconstructed on a grid of pixels, in our case with $100\times 100$ pixels. Since we usually reconstruct astrophysical sources, which typically show quite smooth surface-brightness distributions, a regularization term was applied to penalize very irregular surface-brightness reconstructions. The regularization strength was adapted for each source, which also allowed the reconstruction of relatively clumpy spiral galaxies as the SN host. Specifically, we adopted the curvature form of the regularization term implemented in \GLEE, and refer the reader to \citet{suyu06a} for further details on the definition and procedure.

Furthermore, in each optimization step, the pixel grid of the reconstructed source was mapped back into the image plane, predicting the lensed SN host's intensity distribution. By comparing, for each pixel, the predicted flux values, $\boldsymbol{d}^\text{pred}$, a vector of length $N_\text{pix}$, with the observed ones, $\boldsymbol{d}^\text{obs}$, a vector of the same length, we were able to define a chi-square term as

\be
\chi^2_\text{ext} = (\boldsymbol{d^\text{obs}}-\boldsymbol{d^\text{pred}})^T C_\text{D}^{-1} (\boldsymbol{d^\text{obs}}-\boldsymbol{d^\text{pred}}) \, ,
\ee 

\noindent where $C_\text{D}$ is the image covariance matrix described in \citet{suyu06a}. This led to a very significant increase of the runtime and of the required random access memory for the matrix multiplication. 

Assigning smaller uncertainties to the knots of the SN host instead of a full extended-image mass model as described here is possible and leads to a better reconstruction of the host. However, the extended image model has several further advantages. First, lowering the uncertainties on the multiple images works only for extended sources with significant luminosity structures, such as the SN host galaxy we have here, but not for those with relatively smooth surface-brightness distributions, such as the quasar host reconstructed by \citet{acebron24}, as one cannot identify many knots there. Second, all these knots -- 63 multiple images from 18 different knots in MACS1149 -- need to be properly matched and located, while the pixels used for an extended reconstruction can easily be defined. Third, the lowered uncertainties would only take the identified positions into account, while with our reconstruction we simultaneously made use of the pixel intensities, which have to match each other on a large number of pixels, affecting the deflection angle and magnification values in that region. Fourth, it enables studies of the background source and allows the measurement of extracted quantities such as the half-light radius of the reconstructed source \citep{acebron24}.

Furthermore, this approach can easily be extended to simultaneously model multiband observations of the same source, or to sources at different redshifts. For instance, \citet{schuldt19} presented mass models of the galaxy-scale system called the Cosmic Horseshoe \citep{belokurov07} by modeling the main arc in multiple filters and the radial arc simultaneously.

We then optimized the mass parameter values using simulated annealing and Markov chain Monte Carlo techniques \citep{dunkley05, foreman-mackey13}, as well as emcee \citep{foreman-mackey13}, by minimizing the total chi-square value, $\chi^2= \chi^2_\text{img} + \chi^2_\text{ext}$. However, since the extended image provided almost three orders of magnitude more observables at a specific redshift in our case (the SN host redshift $z_\text{SN}=1.489$) than the point-like images, we tested different weightings of the two chi-square terms. This was achieved by a constant scaling of the combined error map affecting $\chi^2_\text{ext}$ and a constant scaling of the image position uncertainties influencing the $\chi^2_\text{img}$ term. We remark that the precise weighting should depend on the specific scientific case to address. Raising the contribution of the multiple images from sources at different redshifts helps to break the mass-sheet degeneracy \citep{falco85, schneider13}, while more weight on the extended images leads to a better reconstruction of the SN host. Since it is not obvious which weighting is preferred for TDC and its impact on the reconstructed mass model parameters and the quantities derived from them, as for the time delays, we provide an extensive comparison of the different weightings in Sect.~\ref{sec:results}.

In detail, as already noted in Sect.~\ref{sec:model:ImPos}, we considered models  with (\Sw and \Ew ) and without (\Swo and \Ewo) the multiple images of the SN Refsdal and its host, as these image positions were measured from the \HST images, similarly to the flux values used in the extended-image model. We also explicitly note that the SN image positions were excluded in the \Swo and \Ewo models. Furthermore, since we adjusted all the uncertainties to obtain a reduced $\chi^2 \sim 1$ for the correct estimate of uncertainties of the model parameters, we considered, in both cases, two different weightings of the $\chi^2$ terms. For the models \Ew and \Ewo, we fixed the image-position uncertainties to those from \citetalias{schuldt24a}, which means that the $\chi^2_\text{img}$ contribution is negligible as long as the image multiplicity is correct. In case of a wrong image multiplicity, $\chi^2_\text{img}$ and consequently $\chi^2$ are both infinity, and the model is thus rejected, regardless of the extended-image reconstruction (i.e., irrespective of the  $\chi^2_\text{ext}$ value). In other words, in these two models we gave full weight to the extended image, which is the preferred choice when one is particularly interested in studying the source properties. Additionally, we considered the models \Sw and \Swo, where we gave a similar weight ($\sim$50\%) to the two $\chi^2$ terms by adjusting the error map and the image position uncertainties. In these cases, we gave more weight to the sources at different redshifts than in the previous two models, which was crucial for breaking the mass-sheet degeneracy. After obtaining a good fit for each of the four models considered, we ran final emcee chains for 96 days each with ten cores, leading to final chain lengths of more than 600,000 iterations per model. Due to the long runtime, we decided to use an approximate point spread function, that, according to our tests with shorter chains and a point spread function properly constructed from stars in the field, led to the same conclusions overall. The absolute inferred mass parameter values and the details of the source reconstruction (irrelevant in this work) differed only slightly.

\FloatBarrier
\section{Modeling results and comparison to the point-like model}
\label{sec:results}

As described above, we built four different strong lensing models of the MACS1149 cluster, taking into account the extended surface-brightness distribution of the SN host, and compared them to the state-of-the-art point-like model presented by \citetalias{schuldt24a}. An overview of all models is presented in Table~\ref{tab:overview}, where we give the corresponding $\chi^2$ values of the optimized mass models. Since the difference in the models is the weighting of the different $\chi^2$ terms, we report the values for the image positions ($\chi^2_\text{img}$) and extended image ($\chi^2_\text{ext}$) separately. Furthermore, because the associated uncertainties differ for the different models (see Sect.~\ref{sec:model}), we also report chi-square values calculated adopting the same uncertainties, namely the uncertainties from \citetalias{schuldt24a} for the multiple images, resulting in $\chi^2_\text{img,S24}$, and the original error map for the extended image obtained from Eq.~\ref{eq:errmap}, resulting in $\chi^2_\text{ext,orig}$, as these values can be properly compared across the five models.

By looking at the $\chi^2_\text{ext,orig}$ values, we find an expected significant improvement in the reconstructed image, when compared to the point-like model, and we can see that the models \Ew and \Ewo models outperform the \Sw and \Swo models. Moreover, it is also expected that the $\chi^2_\text{img,S24}$ value increases as we give less weight to this term. However, considering that the \Ew and \Ewo models have full weight on the extended image and that the multiple image positions of the SN host and other sources at different redshifts contribute a negligible amount, it is noteworthy that these image positions are nonetheless very well reconstructed. This is shown in the RMS values that increased by only 0\farcs05 for the \Sw and \Swo models and by a factor of approximately two for \Ew and \Ewo, compared to the point-like model, which has its full weight on the multiple image positions. We note that the reported RMS values were always calculated including the image positions of the SN host (labelled RMS(w) in Table~\ref{tab:overview}) to allow a direct and fair comparison between the different models. We remind the reader again that we kept the mass model parametrization exactly as in \citetalias{schuldt24a} to allow a fair comparison to that model, despite the high number of newly introduced observables, so a slightly higher RMS is expected. However, in a future work more flexible mass models could be explored to better reproduce the now incorporated 77,000 \HST pixel values. Moreover, we note that an RMS value of 0\farcs73, which is the highest value we obtained with the model flexibility of \citetalias{schuldt24a}, is comparable to state-of-the-art point-like models from other clusters \citep[e.g.,][]{bergamini19, richard21, furtak23}, where the full weight is on the multiple image positions.

\begin{figure*}[ht!]
    \centering
    \includegraphics[trim={0 0 0 0},clip, width=0.9\linewidth]{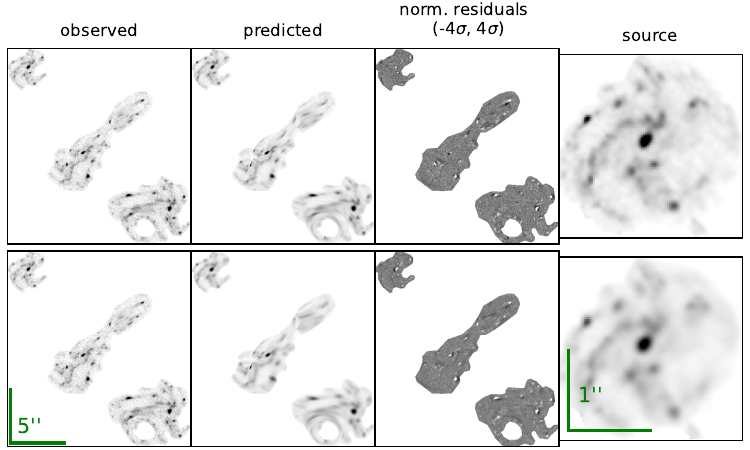}
    \caption{Surface-brightness reconstruction from model \Ew (top row), showing the best reconstruction, and from the point-like model \citetalias{schuldt24a} (bottom row), which was not optimized on the extended image. We show, from left to right, the observed image, the model-predicted image, the normalized residuals, and the reconstructed source (the SN host galaxy). We illustrate the sizes of the images and of the reconstructed sources with green bars of 5\arcsec and 1\arcsec, respectively. We again note that the \Ew model is optimized with an approximate point spread function (see Sect.~\ref{sec:model:ext} for details), which slightly affects the final appearances of the reconstructed image and source.}
    \label{fig:reconstructed}
\end{figure*}

\begin{figure}[t!]
    \centering
    \includegraphics[trim={0 0 0 0},clip, width=\linewidth]{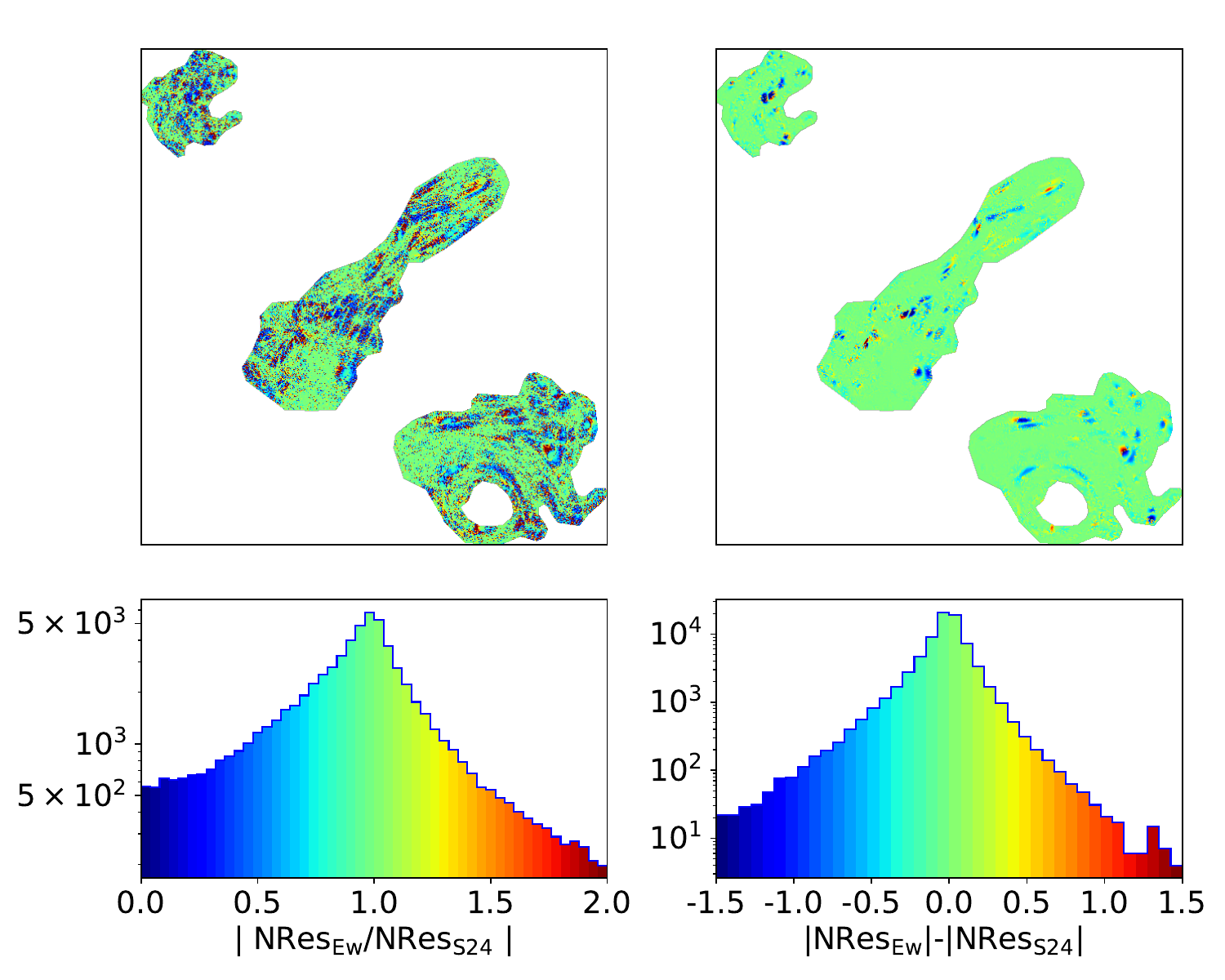}
    \caption{Comparison of normalized residuals (NRes) of the model \Ew and the point-like model by \citetalias{schuldt24a}, as an absolute ratio (left) and absolute difference (right). The bottom panels show the 1D histograms, and the top panels show the corresponding 2D image color-coded in the same way as the histograms.}
    \label{fig:dif_img}
\end{figure}

The improvement in the image reconstruction is visualized in Fig.~\ref{fig:reconstructed}, where we compare \Ew, resulting in the lowest $\chi^2_\text{ext,orig}$ value, with the point-like model \citetalias{schuldt24a}. Although the \Ew  model still shows some residuals, the reconstructed source has significantly more structure than that from the point-like model. We highlight the refinement going from \citetalias{schuldt24a} to \Ew in Fig.~\ref{fig:dif_img}, where we show the corresponding normalized residuals as absolute ratio (left) and difference (right). While the majority of the 77,000 pixels show similar normalized residuals (green), we see a clear trend toward lower values (blue) for the \Ew model, corresponding to a better image reconstruction. It is remarkable that the normalized residuals of more than 200 pixels improved by more than $1.5 \sigma$ compared to the reconstruction with the point-like model, while the normalized residuals for only four pixels increased by $1.5 \sigma$. This reflects the power of the extended image model and translates into a refined cluster-mass reconstruction around the SN host, which is crucial for TDC and studies of the physical properties of the source galaxy.

\begin{figure*}[ht!]
    \centering
    \includegraphics[trim={0 0 0 0},clip, width=0.48\linewidth]{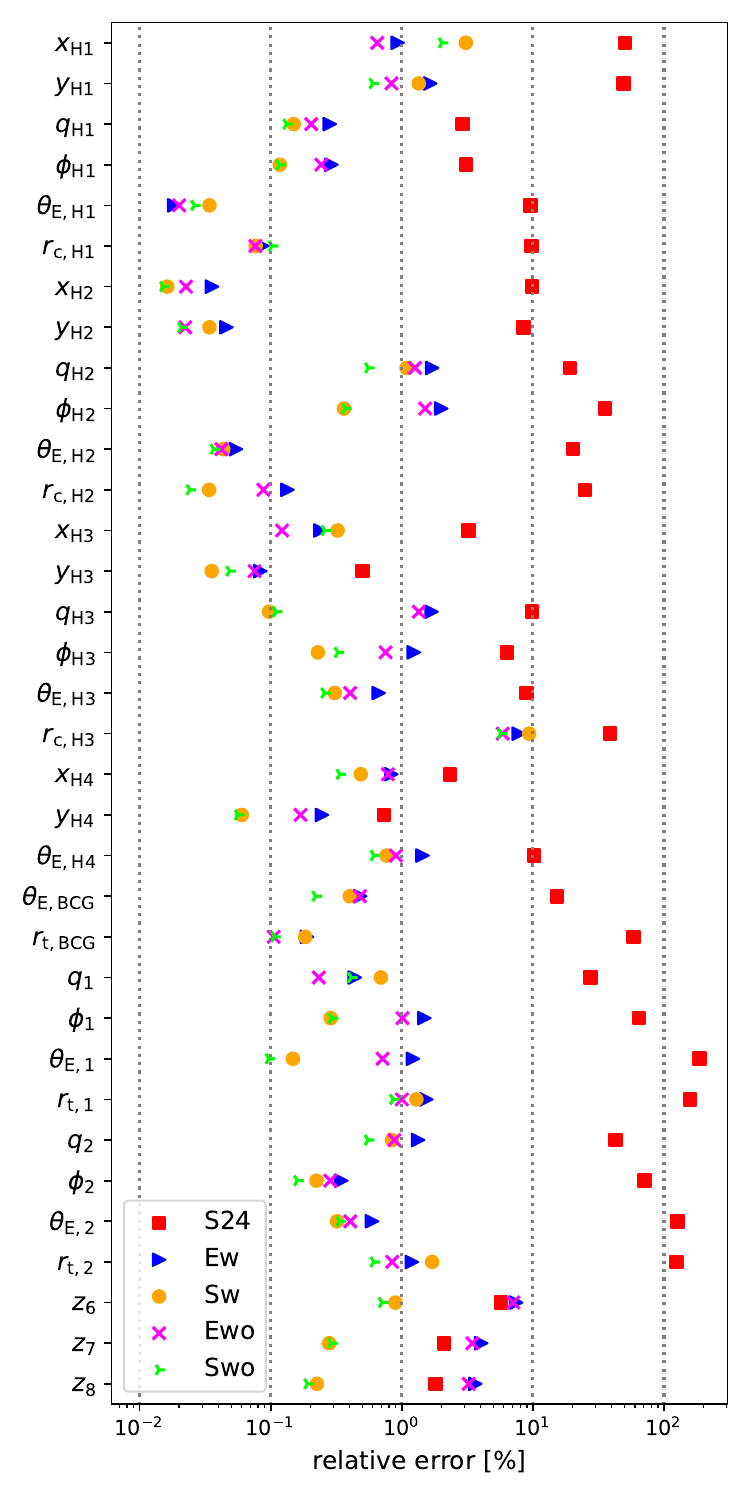} \includegraphics[trim={0 0 0 0},clip, width=0.48\linewidth]{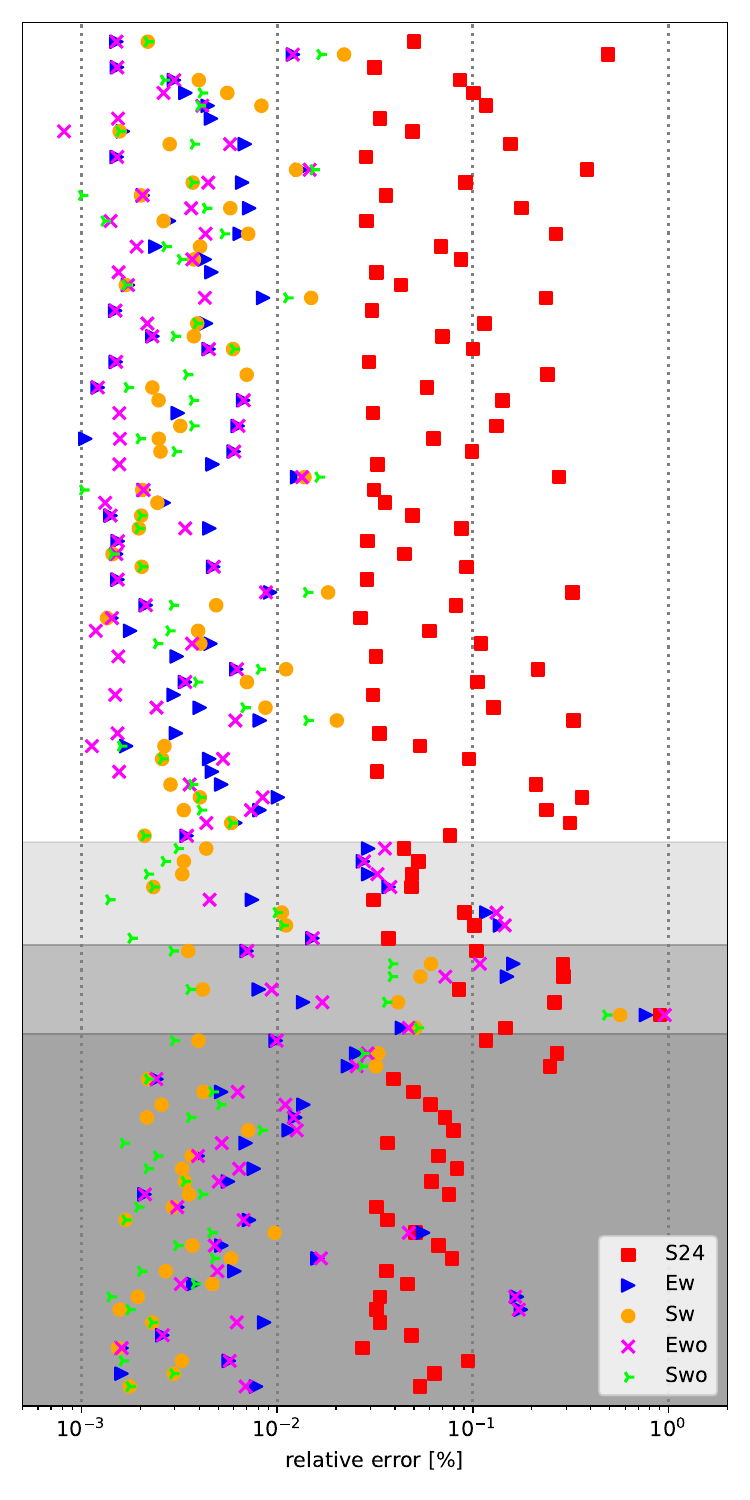}
    \caption{Relative errors of the 34 free parameters (left) and of the magnification factors at the model-predicted positions of the point-like multiple images (right) from the five different models (indicated by different colors and symbols). For the definition of the individual model parameters, see Sects.~\ref{sec:model:parameterization} and \citetalias{schuldt24a}. While the 106 multiple images are, for simplicity, not labeled, we highlight those belonging to the SN and its host (and thus excluded in models \Ewo and \Swo; white background), those with variable redshift (since no spectroscopic redshift is known; light gray), those belonging to the northern galaxy group (see Fig. 2 of \citetalias{schuldt24a}; medium gray), and those from additional sources in the cluster core (dark gray). We observe a significant reduction in the statistical uncertainties of the model parameters (left) and magnifications (right) when we include the 77,000 \HST pixels as direct constraints in the strong lensing model.}
    \label{fig:relerr}
\end{figure*}

By adding the surface-brightness distribution of the SN host galaxy to the other point-like multiple images, we obtain both a significant improvement in the extended image reconstruction and unprecedented precision in the inferred values of the individual mass model parameters. This is illustrated in Fig.~\ref{fig:relerr}, where we show the relative errors of the 34 free parameters for the five models, \Ew, \Ewo, \Sw, \Swo, and the point-like model. The statistical uncertainties of all model parameters, which were estimated by extracting 100 random iterations from the final sampling chains, are reduced by between one and two orders of magnitude. Solely the uncertainties on the three varying redshifts of sources 6, 7, and 8 (see \citetalias{grillo16} and \citetalias{schuldt24a}) are comparable among models S24, \Ew, and \Ewo. This can be explained by the fact that the adopted image-position uncertainty is the same for these three models, while for models \Sw and \Swo it is much lower (see Tab.~\ref{tab:overview}). As a consequence, a small change of the redshifts in the \Sw and \Swo models results in a high $\chi^2_\text{img}$ value, such that these models were penalized. On the other hand, the redshift values of models \Ew and \Ewo can vary more broadly, similarly to those in S24. We note that the relative error is, for some parameters, slightly misleading. For instance, the center of the first DM halo is very close to the origin of the selected coordinate system (see \citetalias{grillo16} and \citetalias{schuldt24a}), so its relative error always appears relatively high. However, since this holds for all five models, the comparison between the models, which is the aim of the plot, is fair and clearly demonstrates that we obtain a significant improvement in precision, regardless of the chosen weighting and inclusion or exclusion of the multiple images of the SN host.

\section{Predicted cumulative total mass profile, time delays, magnifications, and multiband source reconstruction}
\label{sec:td_mass}

\begin{figure}[t!]
    \centering
    \includegraphics[trim={0 0 0 0},clip, width=\linewidth]{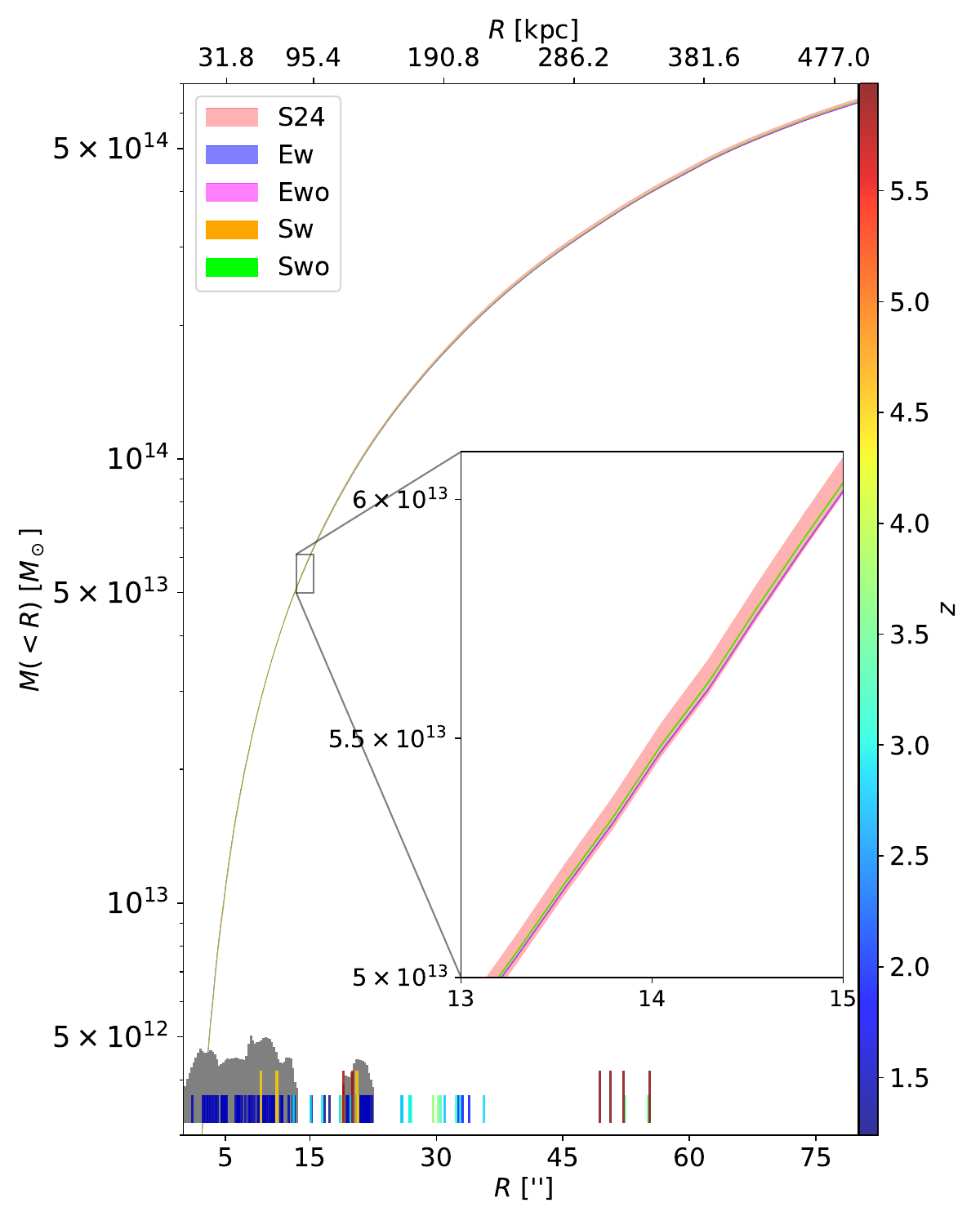}
    \caption{Cumulative projected total mass profile $M(<R)$ of MACS1149 as function of the radius, $R$, measured from the brightest cluster galaxy \citepalias[see][]{grillo16}. The plotted bands correspond to the $1\sigma$ intervals obtained from 100 random iterations of the final sampling chains. We compared our four new extended image models with the point-like model from \citetalias{schuldt24a} and found significantly smaller statistical uncertainties for the extended image models. The positions of the multiple images are marked with small dashes on the lower $x$-axis and color-coded according to their redshifts. The bar lengths reflect the positional uncertainties (a small dash corresponds to an \HST identification, while a long dash corresponds to a \MUSE-only identification; see \citetalias{schuldt24a} for details). We further show the positions of the 77,000 \HST image pixels used in the extended image models with a gray histogram, where the height (in log-scale) corresponds to the number of pixels per radial bin.}
    \label{fig:massprofile}
\end{figure}

After presenting the improvement on the mass parameter values, we considered the impact on some derived quantities. First, we computed the cumulative total mass profile of the cluster MACS1149, as shown in Fig.~\ref{fig:massprofile}. We compared our four new extended-image and previous point-like models and found good agreement among them. We obtained a projected total mass integrated within 500 kpc of $\left( 6.31\pm 0.07 \right) \times 10^{14} M_\odot$, with the $1\sigma$ errors approximately one order of magnitude smaller when the extended surface brightness is included in the models. We note that the quoted uncertainties are only the statistical uncertainties, while the systematic uncertainties should be assessed with different cluster's total mass parametrizations. 


\begin{figure*}[ht!]
    \centering
    \includegraphics[trim={0 0 0 0},clip, width=\linewidth]{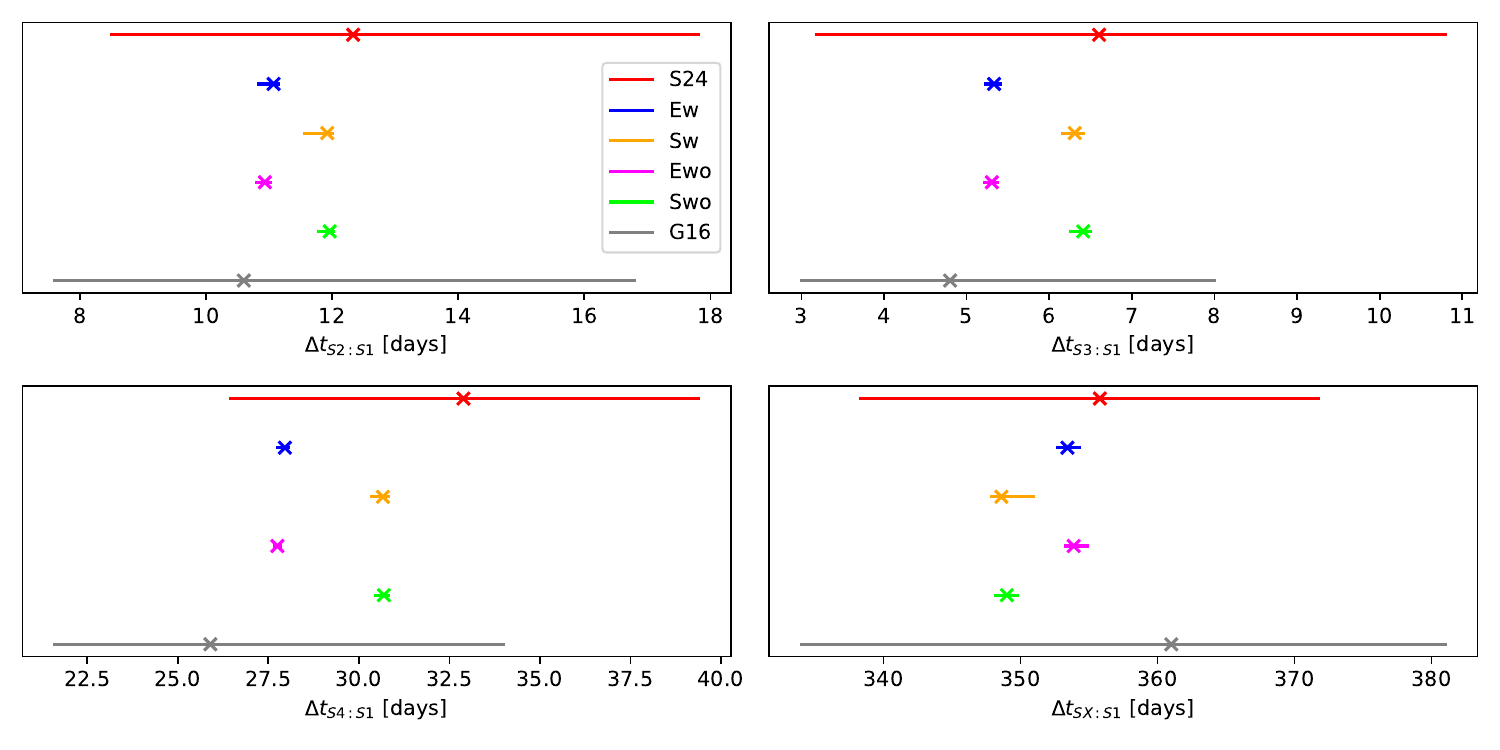}
    \caption{Median values and $1\sigma$ uncertainties of time delays obtained at the model-predicted  positions of the multiple images of the SN with fixed (flat $\Lambda$CDM) cosmology. The four new extended image models and the two previous point-like models by \citetalias{schuldt24a} and \citetalias{grillo16} are shown with different colors. We notice the significantly lower statistical uncertainties for the extended image models compared to the point-like models. We note again that the extended image models are constructed with an approximate point spread function (see Sect.~\ref{sec:model:ext} for details), which slightly affects the exact values of the predicted time delays.}
    \label{fig:TD}
\end{figure*}

Second, we compared the model-predicted values of the time delays between multiple images of the SN observed in MACS1149. Because all our models were obtained with a fixed cosmology (i.e., flat $\Lambda$CDM; see Sect.~\ref{sec:intro}), we only considered the four presented models (\Ew, \Ewo, \Sw, and \Swo) and the point-like models from \citetalias{schuldt24a} and \citetalias{grillo16}, which were built under the same cosmological assumptions. Figure \ref{fig:TD} shows the median values and $1\sigma$ errors of the four predicted time delays for all six models, obtained from 100 different iterations of the final sampling chains. A comparison with the measured time delays would not be fair, given the fixed values of the cosmological parameters in our lensing models. The updated measurement of the values of the cosmological parameters, including the Hubble constant, by exploiting the extended image models exceeds the scope of this study and is thus left for a future work. 

Although the predicted values of the point-like models \citetalias{schuldt24a} and \citetalias{grillo16} differ slightly, they are consistent, given the 1$\sigma$ uncertainties. The uncertainty on $\Delta t_\text{SX:S1}$, which is the longest time delay and thus the most relevant one for cosmological applications, obtained with the mass model from \citetalias{schuldt24a} than with that from \citetalias{grillo16}. This can probably be ascribed to the inclusion of the SX observed position in \citetalias{schuldt24a}, which is different from \citetalias{grillo16}. The latter mass model was developed before the detection of the SN at the SX position (i.e., this position was a blind prediction of that model). Considering this, the predicted value and error of the $\Delta t_\text{SX:S1}$ time delay with model \citetalias{grillo16} represent a remarkable result.

The model-predicted time-delay median values of the extended image models are all well within the 1$\sigma$ intervals of \citetalias{schuldt24a} and \citetalias{grillo16}, and, regardless of the weighting, the obtained statistical uncertainties are more than an order of magnitude smaller. Although a lower uncertainty is expected, given the improved reconstruction of the cluster's total mass (see Fig.~\ref{fig:relerr}), the observed effect is remarkable. The remarkably small error bars are responsible for the fact that only models \Ew and \Ewo, and \Sw and \Swo, respectively, are consistent within $1\sigma$. Consequently, the model-predicted time delays mildly depend on the weighting, i.e., whether we reconstruct the SN host's surface brightness, and thus the local cluster's total mass distribution, in the best way (models \Ew and \Ewo), or if we give 50\% of the weight to the other multiple images, which are at different redshifts, to reduce the mass-sheet degeneracy. We remind the reader that the quoted uncertainties only refer to the statistical errors, and the small differences among the four models can be used to quantify a first source of systematic uncertainties. Until now, with point-like models, the systematic uncertainties were found to be significantly smaller than the statistical ones \citep{grillo20}. Given the increased precision with the source surface-brightness reconstruction, systematic effects will become relevant. 

It is noteworthy that the statistical uncertainty on the time-delay value $\Delta t_\text{SX:S1}$ predicted with point-like models is $\sim$5\%, exceeding the error on the measured time delay, which is on the order of 1.5\% \citep{kelly23b}. Both these terms contribute with similar weights to the resulting precision on the inferred value of $H_0$. Thanks to the extended image models, we achieved a sub-percent statistical uncertainty on the model-predicted time delays. This translates into significantly more precise measurements of the values of the cosmological parameters.

\begin{figure}[ht!]
    \centering
    \includegraphics[trim={0 0 0 0},clip, width=0.8\linewidth]{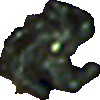}
    \caption{Color composite image of the SN host reconstruction from the Hubble Frontier Field \HST bands F814W (red), F606W (green), and F435W (blue), using the \Ew model. The mapped region was slightly modified compared to that used in the model optimization (shown in Fig.~\ref{fig:reconstructed}) to minimize light contamination from some cluster members and the SN images. This figure underscores our ability to directly model the SN host in multiple bands. The image has an extension of 2\farcs05 and 2\farcs27 along the $x$ and $y$ axes, which correspond to 17.84 kpc and 19.75 kpc, respectively, at a redshift of $z_\text{SN}=1.49$.}
    \label{fig:sr_multiband}
\end{figure}

Another strong application of cluster lensing, including MACS1149 itself \citep[e.g.,][]{stiavelli23, morishita24}, is the study of high-redshift sources that are amplified thanks to the lensing magnification effect. Our extended image models are expected to impact this research field in two ways. As already shown in Fig.~\ref{fig:reconstructed}, we obtained a delensed version of the SN host galaxy. However, while we used a difference image from two \HST bands to minimize the light contribution from all cluster members for the modeling, we can also use a given extended image model to delens the images in a single band. As a demonstration, we used the final \Ew model, achieving the lowest $\chi^2_\text{ext,orig}$, to reconstruct the delensed SN host galaxy from the individual \HST images. In detail, we obtained the reconstruction in three \HST bands, F814W, F606W, and F435W, and in Fig.~\ref{fig:sr_multiband} we show them as a color image. We modified the mapped pixels slightly, mainly the regions of the four SN images, to exclude the light contribution from nearby cluster members and the SN images (slightly visible in F435W). While the extended image modeling is computationally very expensive, the source reconstruction is not (this is  roughly equivalent to a single model iteration), and thus it can be easily done in multiple bands and with images from other telescopes.

The extended image modeling and the delensing of background sources is only possible for extended ones, such as the SN host with 77,000 \HST pixels. This technique is not directly applicable for compact objects, extending only over a small amount of pixels. Therefore, resulting in an additional application scenario, we show the predicted magnifications of the 106 multiple images, which is a crucial measurement for studying background sources, in Fig.~\ref{fig:relerr} \citep[see][and \citetalias{schuldt24a} for the exact coordinates]{grillo16}. As we see from Fig.~\ref{fig:relerr} (right panel), where we show the relative uncertainties on the magnification, we found an improvement by $\sim$1 order of magnitude between the classical point-like model and the four extended image models. As expected, this holds specifically for the multiple images from the SN host (white background in Fig.~\ref{fig:relerr}). Therefore, we expect this model approach to impact the high-redshift studies in general. It is understandable that models \Ew and \Ewo preform better than \Sw and \Swo on the multiple image positions of the SN host, while the performance on the other images is overall worse. This holds in particular for the images with photometric redshifts (light gray band in Fig.~\ref{fig:relerr}), as these images add very little information to \Ew and \Ewo, while, because of the reduced image position uncertainty, they contribute significantly to models \Sw and \Swo. 

It is also interesting to see the behavior of the northern galaxy group, where we lack a significant improvement for most of the seven multiple images (medium gray background in Fig.~\ref{fig:relerr}). This is likely due to the angular distance to the SN host galaxy, where the additional observables are located. In particular, models \Ew and \Ewo lack any significant constraints in this region, which underscores that the choice of weighting depends on the project's scope and can be used to mitigate this effect. However, we did not find a strong dependency on the weighting for the overall predicted quantities (see Figs.~\ref{fig:massprofile} and \ref{fig:TD}). Furthermore, we explicitly note that all extracted quantities (total cluster mass, time delays, etc.) fall well within the previous statistical uncertainties from our point-like mass model \citetalias{schuldt24a}, while the differences between our new mass models reflect systematic uncertainties that became relevant. These systematic uncertainties are well below the statistical uncertainties of the mass model from \citetalias{schuldt24a} (see Fig.~\ref{fig:TD}), which is in agreement with previous findings by \citet{treu16} and \citet{grillo20}, for example.

\section{Summary and conclusion}
\label{sec:conclusion}

We obtained the first total mass model of the strong lensing cluster MACS1149 by combining 106 identified multiple images and the extended surface-brightness distribution of the SN host galaxy. The cluster mass was modeled with parameterized profiles, following the point-like model from \citetalias{schuldt24a} to enable a direct comparison, and composed of three elliptical DM halos, a spherical DM halo associated with the galaxy group in the north (see Fig. 2 in \citetalias{schuldt24a}), two elliptical profiles for two cluster members close to the SN host, and a spherical profile for each of the 306 additional cluster members, where their Einstein and truncation radii were scaled according to given scaling relations. Additionally, three background sources without spectroscopic redshift were included with a variable redshift value. In total, we optimized 34 free parameters (see Fig.~\ref{fig:relerr}). While this seems a small number of parameters, including a few hundred mass components and several dozen background sources already leads to a relatively long runtime, when optimizing a model on the image plane. We faced significant computational challenges by incorporating 77,000 \HST pixels in the modeling. Furthermore, the random access memory (RAM) increased remarkably, from $\sim$1GB/core to around 180GB/core, making a parallelized optimization extremely difficult. 

Since we considered the flux values of 77,000 \HST pixels in addition to the 212 multiple-image positions ($x$ and $y$ coordinates for 106 images) as observables, we examined, for the first time, different combinations of the two groups of observables. Namely, we tested the effect of the weighting with four different models, fixing the model parametrization as in \citetalias{schuldt24a}. In detail, we analyzed a first version effectively with full weight on the extended image of the SN host, and a second version where we rescaled the image position uncertainties and the surface-brightness error map so that both terms contributed similarly to the total chi-square value. Furthermore, for the two previous cases, we built a version with all identified image positions, and one excluding those of the SN host, as these were considered as already included in the \HST pixels. Regardless of the weighting and of the included or excluded SN host image positions, the values of the model parameters were measured with precision between one and two orders of magnitude higher than in the point-like model (see Fig.~\ref{fig:relerr}). We also found slightly lower statistical uncertainties for models with equal weights (i.e., models \Sw and \Swo). As expected, we obtained a better reconstruction of the SN host with the models adopting full weight on the extended images (see Fig.~\ref{fig:dif_img} and Tab.~\ref{tab:overview}). 

Going beyond the simple comparison of the parameter values, we quantified the impact on some derived quantities of the models. First, given the powerful use of gravitational lensing to measure the total mass of a lens, we obtained the projected cumulative total mass profile of the lensing cluster MACS1149 (see Fig.~\ref{fig:massprofile}) with remarkably small errors. Then, since the cluster produced multiple images of SN Refsdal, we predicted the values of their time delays and compared them with those from \citet{grillo16} and \citetalias{schuldt24a} within a fixed cosmological model. In particular, we did not find any bias on the predicted time delays using the different weighting and showed that the statistical uncertainty, associated with the remaining uncertainty in the cluster's total mass reconstruction, decreased from $\sim$5\%, for the point-like model, to less than 1\% for the extended image models. This will translate into significantly more precise estimates of the values of the cosmological parameters. We thus reached a regime in which the systematic errors need to be quantified, as we can see from the minor differences between the four different mass model predictions. Moreover, we measured the magnification factors for all 106 identified multiple images with the four new mass models and the point-like model \citetalias{schuldt24a} and compared their relative errors (see Fig.~\ref{fig:relerr}). In all cases, we found a significant improvement in precision when taking the surface-brightness distribution of the SN host into account. 

Furthermore, we reconstructed the SN host galaxy in multiple \HST bands and show, for illustration purposes, a color-composite image in Fig.~\ref{fig:sr_multiband}. Thanks to extended image models, such as those developed in this work, precise source reconstructions will be possible in the future, using the  \HST, \textit{James Webb} Space Telescope (JWST), or other observations. This will likely significantly impact high-redshift source studies. While it is clear that for this specific science case the best source reconstruction is preferred (i.e., model \Ew or \Ewo), the other weighting (i.e., model \Sw and \Swo) is likely more effective in reducing the mass-sheet degeneracy thanks to the multiple images at different redshifts, and it is thus more suitable for cosmological applications. This shows that the preferred weighting might depend on the specific scientific goal.

In conclusion, we present the potentialities of extended strong lensing models in the Hubble Frontier Field galaxy cluster MACS1149, showing the first multiple-imaged and spatially resolved supernova. We demonstrate the impact on several scientific applications, from TDC to high-redshift studies. With an improvement of more than one order of magnitude in the statistical uncertainties of the model parameters, this work paves the way for a new generation of mass models of cluster lenses exploiting high-resolution imaging data from the \HST, JWST, and Euclid telescope. Accelerated lensing codes such as Herculens \citep{galan22_Herculens}, Gravity.jl \citep{lombardi24}, or GPU-based \GLEE and \GLaD \citep{wang25} will play a fundamental role in reducing the computational time and thus making analyses of this kind feasible.

\begin{acknowledgements}
We thank the referee for constructive comments that helped to improve the clarity of the presented work.
SS has received funding from the European Union’s Horizon 2022 research and innovation programme under the Marie Skłodowska-Curie grant agreement No 101105167 — FASTIDIoUS. We acknowledge financial support through grants PRIN-MIUR 2017WSCC32 and 2020SKSTHZ. PB acknowledges financial support through grant PRIN-MIUR 2020SKSTHZ and support from the Italian Space Agency (ASI) through contract ``Euclid - Phase E'', INAF Grants ``The Big-Data era of cluster lensing'' and ``Probing Dark Matter and Galaxy Formation in Galaxy Clusters through Strong Gravitational Lensing''.
This work is supported in part by the Deutsche Forschungsgemeinschaft (DFG, German Research Foundation) under Germany's Excellence Strategy -- EXC-2094 -- 390783311.  
This work uses the following software packages:
\href{https://github.com/astropy/astropy}{\texttt{Astropy}}
\citep{astropy1, astropy2},
\href{https://github.com/dfm/emcee}{\texttt{Emcee}}
\citep{foreman-mackey13},
\GLEE
\citep{suyu10a_GLEE, suyu12b_GLEE},
\gleetoolspy
\citep{schuldt23b},
\href{https://github.com/matplotlib/matplotlib}{\texttt{matplotlib}}
\citep{matplotlib},
\href{https://github.com/numpy/numpy}{\texttt{NumPy}}
\citep{numpy1, numpy2},
\href{https://www.python.org/}{\texttt{Python}}
\citep{python},
\href{https://github.com/scipy/scipy}{\texttt{Scipy}}
\citep{scipy}.
\end{acknowledgements}

\bibliographystyle{aa}
\bibliography{aa57680-25}

\end{document}